\documentclass[twocolumn,longbibliography,superscriptaddress,showkeys]{revtex4-1}
\pdfoutput=1

\usepackage{graphicx}
\usepackage{dcolumn}
\usepackage{bm}
\usepackage{color}
\usepackage{url}
\usepackage{multirow} 
\usepackage{xcolor}
\usepackage{epsfig}
\usepackage{amsmath}
\usepackage{amssymb}
\usepackage{longtable}
\usepackage{float}
\usepackage{mathtools}
\usepackage{braket}
\usepackage[colorlinks]{hyperref}
\usepackage{xparse}
\NewDocumentCommand{\ceil}{s O{} m}{%
  \IfBooleanTF{#1} 
    {\left\lceil#3\right\rceil} 
    {#2\lceil#3#2\rceil} 
}
\newcommand{\be}{\begin{equation}}
\newcommand{\ee}{\end{equation}}
\newcommand{\bd}{\begin{displaymath}}
\newcommand{\ed}{\end{displaymath}}
\newcommand{\BE}{\begin{eqnarray}}
\newcommand{\EE}{\end{eqnarray}}

\newcommand{\bx}{\ensuremath{\mathbf{x}}}

\newcommand{\tr}{\mathrm{Tr}}

\providecommand{\abs}[1]{\lvert#1\rvert}

\begin{document}

\title{A generative modeling approach for benchmarking and training shallow quantum circuits}

\author{Marcello Benedetti}
\affiliation{Department of Computer Science, University College London, WC1E 6BT London, UK}
\affiliation{Cambridge Quantum Computing Limited, CB2 1UB Cambridge, UK}

\author{Delfina Garcia-Pintos}
\affiliation{Qubitera, LLC., Mountain View, CA 94041, USA}

\author{Oscar Perdomo}
\affiliation{Qubitera, LLC., Mountain View, CA 94041, USA}
\affiliation{Rigetti Computing, 2919 Seventh Street, Berkeley, CA 94710-2704, USA}
\affiliation{Department of Mathematics, Central Connecticut State University, New Britain, CT 06050, USA}

\author{\mbox{Vicente Leyton-Ortega}}
\affiliation{Qubitera, LLC., Mountain View, CA 94041, USA}
\affiliation{Rigetti Computing, 2919 Seventh Street, Berkeley, CA 94710-2704, USA}

\author{Yunseong Nam}
\affiliation{IonQ, Inc., College Park, MD 20740, USA}

\author{Alejandro Perdomo-Ortiz}
\email{Correspondence: alejandro@zapatacomputing.com}
\affiliation{Department of Computer Science, University College London, WC1E 6BT London, UK}
\affiliation{Qubitera, LLC., Mountain View, CA 94041, USA}
\affiliation{Rigetti Computing, 2919 Seventh Street, Berkeley, CA 94710-2704, USA}
\affiliation{Quantum Artificial Intelligence Lab., NASA Ames Research Center, Moffett Field, CA 94035, USA}
\affiliation{Zapata Computing Inc., 439 University Avenue, Office 535, Toronto, ON M5G 1Y8, Canada}

\date{June 2, 2019}

\begin{abstract}
Hybrid quantum-classical algorithms provide ways to use noisy intermediate-scale quantum computers for practical applications. Expanding the portfolio of such techniques, we propose a quantum circuit learning algorithm that can be used to assist the characterization of quantum devices and to train shallow circuits for generative tasks. The procedure leverages quantum hardware capabilities to its fullest extent by using native gates and their qubit connectivity. We demonstrate that our approach can learn an optimal preparation of the Greenberger-Horne-Zeilinger states, also known as ``cat states". We further demonstrate that our approach can efficiently prepare approximate representations of coherent thermal states, wave functions that encode Boltzmann probabilities in their amplitudes. Finally, complementing proposals to characterize the power or usefulness of near-term quantum devices, such as IBM's quantum volume, we provide a new hardware-independent metric called the qBAS score. It is based on the performance yield in a specific sampling task on one of the canonical machine learning data sets known as Bars and Stripes. We show how entanglement is a key ingredient in encoding the patterns of this data set; an ideal benchmark for testing hardware starting at four qubits and up. We provide experimental results and evaluation of this metric to probe the trade off between several architectural circuit designs and circuit depths on an ion-trap quantum computer. 
\end{abstract}


\maketitle

\section{Introduction}~\label{s:intro}
What is a good metric for the computational power of noisy intermediate-scale quantum~\cite{Preskill2018} (NISQ) devices? Can machine learning (ML) provide ways to benchmark the power and usefulness of NISQ devices? How can we capture the performance scaling of these devices as a function of circuit depth, gate fidelity, and qubit connectivity? In this work, we design a hybrid quantum-classical framework called data-driven quantum circuit learning (DDQCL) and address these questions through simulations and experiments.

Hybrid quantum-classical algorithms, such as the variational quantum eigensolver~\cite{Peruzzo2014,McLean2016} (VQE) and the quantum approximate optimization algorithm~\cite{Farhi2014,Hadfield2017} (QAOA), provide ways to use NISQ computers for practical applications. For example, VQE is often used in quantum chemistry when searching the ground state of the electronic Hamiltonian of a molecule~\cite{o2016scalable,colless2017robust,kandala2017hardware}. QAOA is used in combinatorial optimization to find approximate solutions of a classical Ising model~\cite{Moll2017} and it has been demonstrated on a 19-qubit device for the task of clustering synthetic data~\cite{otterbach2017unsupervised}. Other successful hybrid approaches based on genetic algorithms were proposed for approximating quantum adders and training quantum \mbox{autoencoders~\cite{Romero2017,lamata2017quantum,li2017approximate}.} In all these examples, there is a clear-cut objective function describing the cost associated with each candidate solution. The task is then to optimize it exploiting both quantum and classical resources. 

Although optimization tasks offer a great niche of applications, probabilistic tasks involving sampling have most of the potential to prove quantum advantage in the near-term~\cite{Farhi2016,PerdomoOrtiz2017,Harrow2017}. For example, learning probabilistic generative models is in many cases an intractable task, and quantum-assisted algorithms have been proposed for both gate-based~\cite{Wiebe-arXiv-2015,kieferova2016tomography} and quantum annealing \mbox{devices~\cite{Benedetti-2016,Amin-arXiv-2016,Benedetti2017,Benedetti2017b,wittek2017quantum}.} Differently from optimization, the learning of a generative model is not always described by a clear-cut objective function. All we are given as input is a data set, and there are several cost functions that could be used as a guide, each one with their own advantages, disadvantages, and assumptions. 

Here we present a hybrid quantum-classical approach for generative modeling on gate-based NISQ devices which heavily relies on sampling. We use the $2^N$ amplitudes of the wave function obtained from a $N$-qubit quantum circuit to construct and capture the correlations observed in a data set. As Born's rule determines the output probabilities, this model belongs to the class of models called Born machines~\cite{Cheng2017}. Previous implementations of Born machines~\cite{stoudenmire2016supervised,han2017unsupervised,liu2017machine,gao2017efficient} often relied on the construction of tensor networks and their efficient manipulation through graphical-processing units. Our work differs in that Born's rule is naturally implemented by a quantum circuit executed on a NISQ hardware. Following the notation of subsequent work~\cite{liu2018differentiable}, we refer to our generative model as a \textit{quantum circuit Born machine}.

By employing the quantum circuit itself as a model for the data set, we also differentiate from quantum algorithms that target specific probability distributions. For example, in Ref.~\cite{Verdon2016} the authors developed a hybrid algorithm to approximately sample from a Boltzmann distribution. The samples are used to update a classical generative model, which requires running the hybrid algorithm for each training iteration. In contrast, our work does not assume a Boltzmann distribution and therefore does not require a specific sampling beyond Born's rule.

Our work is also in contrast with other quantum-assisted ML work (see e.g. Refs.~\cite{Lloyd-NatPhys-2014,kerenidis2016quantum,Brandao2017,schuld2017quantum}) requiring fault-tolerant quantum computers, which are not expected to be readily available in the near-term~\cite{Mohseni2017}. Instead, our circuits are carefully designed to exploit the full power of the underlying NISQ hardware without the need for a compilation step.

On the benchmarking of NISQ devices, quantum volume~\cite{bishop2017quantum,Moll2017} has been proposed as an architecture-neutral metric. It is based on the task of approximating a specific class of random circuits and estimating the associated effective error rate. This is very general and it is indeed useful for estimating the computational power of a quantum computer. In this paper, we propose the \textit{qBAS score}, a complementary metric designed for benchmarking hybrid quantum-classical systems. The score is based on the generative modeling performance on a canonical synthetic data set which is easy to generate, visualize, and validate for sizes up to hundreds of qubits. Yet, implementing a shallow circuit that can uniformly sample such data is hard; we will show that some candidate solutions require large amount of entanglement. Hence, any miscalibration or environmental noise will affect this single performance number, enabling comparison between different devices or across different generations of the same device. Moreover, the score depends on the classical resources, hyper-parameters, and various design choices, making it a good choice for the assessment of the hybrid system as a whole.

Our design choices are based on the setup of existing ion trap architectures. We choose the ion trap because of its full connectivity among qubits~\cite{linke2017experimental} which allows us to study several circuit layouts on the same device. Our experiments are carried out on an ion trap quantum computer hosted at the University of Maryland~\cite{Debnath2016}.

\section{Results}
\subsection*{The learning pipeline}~\label{s:DDVQAlearning}
In this Section, we present a hybrid quantum-classical algorithm for the unsupervised machine learning task of approximating an unknown probability distribution from data. This task is also known as generative modeling. First, we describe the data and the model, then we present a training method.

The data set ${ \mathcal{D} = (\mathbf{x}^{(1)}, \cdots, \mathbf{x}^{(D)}) }$ is a collection of $D$ independent and identically distributed random vectors. The underlying probabilities are unknown and the target is to create a model for such distribution. For simplicity, we restrict our attention to $N$-dimensional binary vectors $\mathbf{x}^{(d)} \in \{-1,+1\}^N$, e.g. black and white images. This gives us an intuitive one-to-one mapping between observation vectors and the computational basis of an \mbox{$N$-qubit} quantum system, that is ${ \mathbf{x} \leftrightarrow \ket{\mathbf{x}} = \ket{x_1 x_2 \cdots x_N} }$. Note that standard binary encodings can be used to implement integer, categorical, and approximate continuous variables.

Provided with the data set $\mathcal{D}$, our goal is now to obtain a good approximation to the target probability distribution $P_{\mathcal{D}}$. A quantum circuit model with fixed depth and gate layout, parametrized by a vector $\boldsymbol{\theta}$, prepares a wave function $\ket{\psi(\boldsymbol{\theta})}$ from which probabilities are obtained according to Born's rule ${ P_{\boldsymbol{\theta}}(\mathbf{x}) = \abs{ \braket{ \mathbf{x} | \psi(\boldsymbol{\theta})}}^2 }$. Following a standard approach from generative machine learning~\cite{Bengio-Book}, we can minimize the Kullback-Leibler (KL) divergence~\cite{kullback1951information} $D_{KL}[P_{\mathcal{D}}|P_{\boldsymbol{\theta}}]$ from the circuit probability distribution in the computational basis $P_{\boldsymbol{\theta}}$ to the target probability distribution $P_{\mathcal{D}}$. Minimization of this quantity is directly related to the minimization of a well known cost function: the negative log-likelihood ${\mathcal{C} (\boldsymbol{\theta}) = - \frac{1}{D}\sum_{d=1}^{D} \ln ( P_{\boldsymbol{\theta}}(\mathbf{x}^{(d)}) )}$. However, there is a caveat; as probabilities are estimated from frequencies of a finite number of measurements, low-amplitude states could lead to incorrect assignements. For example, an estimate $P_{\boldsymbol{\theta}}(\mathbf{x}^{(d)})=0$ for some $\mathbf{x}^{(d)}$ in the data set would lead to infinite cost. To avoid singularities in the cost function, we use a simple variant
\be\label{e:cost_func}
\mathcal{C}_{nll} (\boldsymbol{\theta}) = - \frac{1}{D}\sum_{d=1}^{D} \ln ( \max (\epsilon, P_{\boldsymbol{\theta}}(\mathbf{x}^{(d)}))),
\ee
where $\epsilon>0$ is a small number to be chosen. Note that the number of measurements needed to obtain an unbiased estimate of the relevant probabilities may not scale favorably with the number of qubits $N$. In the Supplementary Material we suggest alternative cost functions such as the moment matching error and the earth mover's distance.

After estimating the cost, we update the parameter vector $\boldsymbol{\theta}$ to further minimize the cost. This can in principle be done by any suitable classical optimizer. We used a gradient-free algorithm called particle swarm optimization (PSO)~\cite{kennedy1995particle,shi1998modified} as previously done in the context of quantum chemistry~\cite{colless2017robust}. The algorithm iterates for a fixed number of steps, or until a local minimum is reached and the cost does not decrease.

We chose the layout of the model circuit to be of the following form. Let us consider a general circuit parametrized by single qubit rotations $\{\theta^{(l,k)}_{i}\}$ and two-qubit entangling rotations $\{\theta^{(l)}_{ij}\}$. The subscripts denote qubits involved in the operation, $l$ denotes the layer number and $k\in\{1,2,3\}$ denotes the rotation identifier. The latter is needed as we decompose an arbitrary single qubit rotation into three simpler rotations (see Section~\ref{ss:simdetails} for further details about some exceptions and potential simplifications depending on the native gates available in each specific hardware). Inspired by the gates readily available in ion trap quantum computers, we use alternating layers of arbitrary single qubit gates (odd layers) and M{\o}lmer-S{\o}rensen $XX$ entangling gates~\cite{Sorensen1999, Sorensen2000, Benhelm2008, Maslov2017} (even layers) as our model. All parameters are initialized at random. 

It is important to note that in our model the number of parameters is fixed and is independent of the size $D$ of the data set. This means we can hope to obtain a good approximation to the target distribution only if the model is flexible enough to capture its complexity. Increasing the number of layers or changing the topology of the entangling layer alter this flexibility, potentially improving the quality of the approximation. However, we anticipate that such flexible models are more challenging to optimize because of their larger number of parameters. Principled ways to choose the circuit layout and to regularize its parameters could significantly help in such a case.

\begin{figure}[t]
\centering
\includegraphics[width=.48\textwidth]{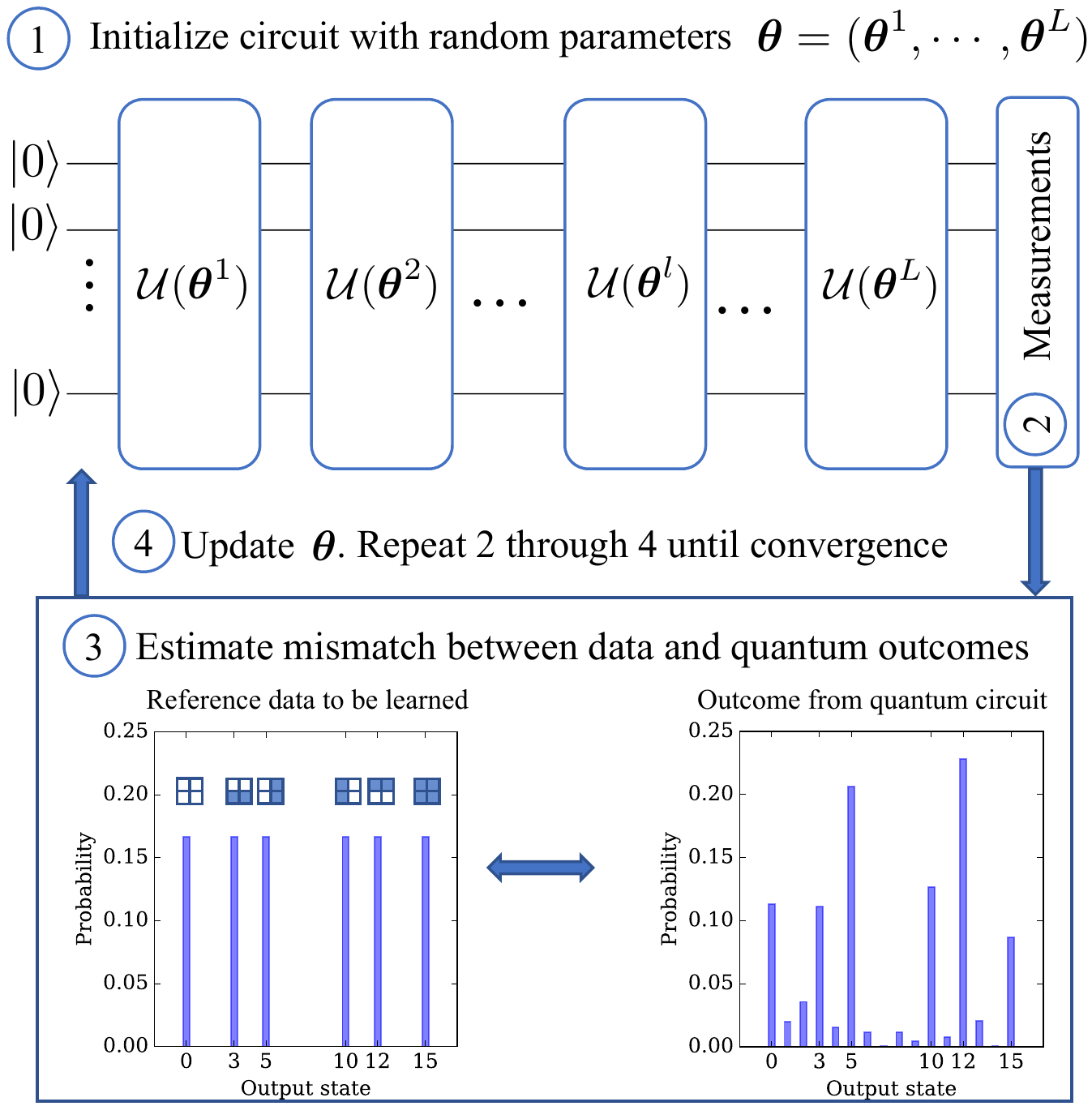}
\caption{{\it General framework for data-driven quantum circuit learning (DDQCL).} Data vectors are interpreted as representative samples from an unknown probability distribution and the task is to model such distribution. The $2^N$ amplitudes of the wave function resulting from an $N$-qubit quantum circuit are used to capture the correlations observed in the data. Training of the quantum circuit is achieved by successive updates of the parameters $\boldsymbol{\theta}$, corresponding to specifications of single qubit operations and entangling gates. In this work, we use arbitrary single qubit rotations for the odd layers, and M{\o}lmer-S{\o}rensen $XX$ gates for the even layers. At each iteration, measurements from the quantum circuit are collected and contrasted with the data through evaluation of a cost function which tracks the learning progress.}
\label{f:ddvqa}
\end{figure}

As summarized in Figure~\ref{f:ddvqa}, the learning algorithm iteratively adjusts all the parameters to minimize the value of the cost function. At any iteration the user-defined cost is approximated using both samples from the data set and measurements from the quantum hardware, hence the name data-driven quantum circuit learning (DDQCL).

\subsection*{The qBAS score}~\label{s:qbas}
\begin{figure*}
\centering
\includegraphics[width=.98\textwidth]{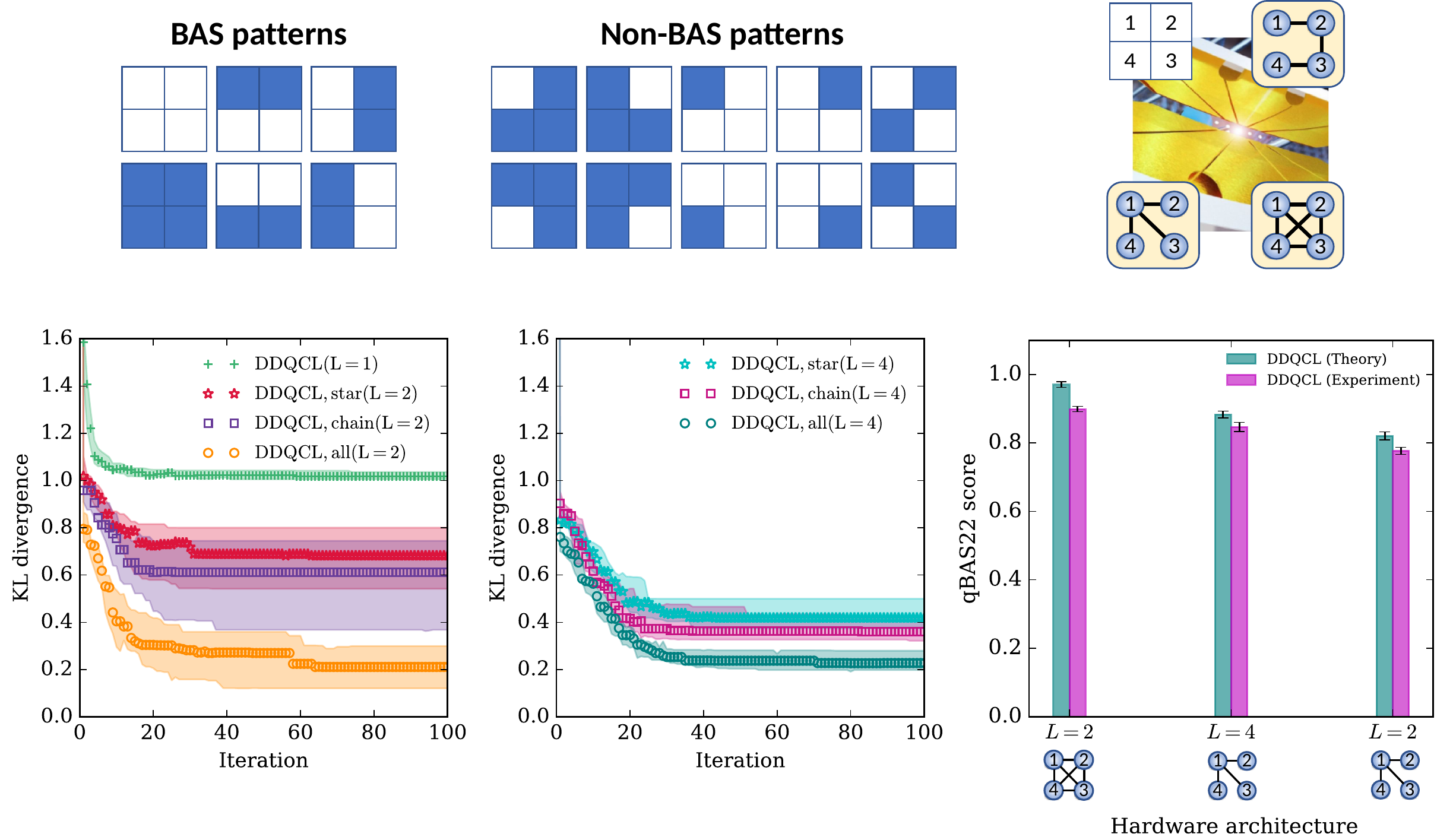}
\caption{\textit{DDQCL on the BAS data set.} The top left panel shows patterns that belong to BAS$(2,2)$ our quantum circuit is to generate. The top central panel shows undesired patterns. On the top right panel, we show a possible mapping of the $4$ pixels to $N=4$ qubits, and we show some of the qubit-to-qubit connectivity topologies that can be set up in entangling layer and natively implemented by the ion trap quantum computer (e.g \textit{chain}, \textit{star}, and \textit{all}). The bottom left panel shows the results of DDQCL simulations of shallow circuits with different topologies. We show the bootstrapped median and $90\%$ confidence interval over the distribution of medians of the KL divergence as learning progresses for $100$ iterations. The mean-field-like circuit $L=1$ (green crosses) severely underperforms. A significant improvement is obtained with $L=2$, where most of the parameters for $XX$ gates have been learned to their maximum entangling value. These observations indicate that entanglement is a key resource for learning the BAS data set. Note that for $L=2$ the choice of topology becomes a key factor for improving the performance. The chain topology (purple squares) performs slightly better than the star topology (red stars) even though they have the same number of parameters. The all-to-all topology (orange circles) significantly outperform all the others as it has more expressive power. The bottom central image extends the previous analysis to deeper circuits with $L=4$ and approximatively twice the number of parameters. All the topologies achieve a lower median KL divergence and the confidence intervals shrink. The bottom right panel shows the bootstrapped mean qBAS$(2,2)$ score and $95\%$ confidence interval for simulations (green bars) and experiments on the ion trap quantum computer hosted at University of Maryland (pink bars).}
\label{f:bas}
\end{figure*}

Bars and stripes (BAS)~\cite{MacKay-book-2002} is a synthetic data set of images that has been widely used to study generative models for unsupervised machine learning. For ${ n \times m }$ pixels, there are ${ N_{{\rm BAS}(n,m)} = 2^n + 2^m - 2 }$ images belonging to BAS, and they can be efficiently produced and visualized. The probability distribution is $1/N_{{\rm BAS}(n,m)}$ for each pattern belonging to BAS$(n,m)$, and zero for any other pattern. Figure~\ref{f:bas} on the top left panel shows patterns belonging to BAS$(2,2)$, while the top central panel shows the remaining patterns.

We use DDQCL to learn a circuit that encodes all the BAS patterns in the wave function of a quantum state. This also allows us to design the qBAS$(n,m)$ score: a task specific figure of merit to assess the performance of shallow quantum circuits. In a single number, it captures the model capacity of the circuit layout and intrinsic hardware strengths and limitations in solving a complex sampling task that requires a fair amount of entanglement. It takes into account the model circuit depth, gate fidelities, and any other architectural design aspects, such as the quantum hardware's qubit-to-qubit connectivity and native set of single and two-qubit gates. It also takes classical resources such as the choice of cost function, optimizer, and hyper-parameters, into account. Therefore, it can be used to benchmark the performance of the components of the hybrid quantum-classical system.

The qBAS$(n,m)$ score is an instantiation of the F$_1$ score widely used in the context of information retrieval. The F$_1$ score is defined as the harmonic mean of the precision $p$ and the recall $r$, i.e. F$_1 = 2 p r/(p+r)$. The precision $p$ indicates the ability to retrieve states which belong to the data set of interest~\footnote{The meaning and usage of precision in the field of information retrieval differs from the definition of precision within other branches of science and statistics.}. In our context this is the number of measurements that belong to the BAS$(n,m)$ data set, divided by the total number of measurements $N_{\rm{reads}}$ performed. The recall $r$ is the capacity of the model to retrieve the whole spectrum of patterns belonging to the desired data set. In our case, if we denote the number of unique patterns that were measured as $d (N_{\rm{reads}})$, then $r = d (N_{\rm{reads}})/N_{{\rm BAS}(n,m)}$. To score high (F$_1$ $\approx 1.0$), both high precision ($p \approx 1.0$) and high recall ($r \approx 1.0$) are required.

The F$_1$ score is a useful measure for the quality of information retrieval and classification algorithms, but for our purposes it has a caveat: the dependence of $r$ on the total number of measurements. As an example, consider a model that generates only BAS patterns, i.e. its precision is 1.0, but with highly heterogeneous distribution. If some of the BAS patterns have infinitesimally small probability, we can still push the recall to 1.0 by taking a large number of measurements $N_{\rm{reads}} \rightarrow \infty$. This is not desirable since our purpose is to evaluate circuits on the task of uniformly sampling all the patterns from BAS$(n,m)$. Therefore, to define a unique score it is important to fix $N_{\rm{reads}}$ to a reasonable value such that $r \approx 1.0$ under the assumption of the model distribution being equal to the target distribution $P_{ {\rm BAS}(n,m)}$, but not so large as to make the score insensitive to deviations from the target distribution. Assuming a perfect model distribution $P_{ {\rm BAS}(n,m)} = 1/N_{{\rm BAS}(n,m)}$, the expected number of measurements to obtain a value of $r=1.0$ can be estimated using the famous \textit{coupon collector's problem}. For our purposes here, we set $N_{\rm{reads}}$ to be equal to the expected number of samples that need to be drawn to collect all the $N_{{\rm BAS}(n,m)}$ patterns (``coupons''). That is, $N_{\rm{reads}} = N_{{\rm BAS}(n,m)} H_{N_{{\rm BAS}(n,m)}}$, where $H_k$ is the $k$-th harmonic number. Computed values of $N_{\rm{reads}}$ are provided in Table~\ref{t:BASnm} for different values of $n$ and $m$ up to 100 qubits. As shown in the table, the number of readouts required to determine qBAS$(n,m)$ are within experimental capabilities of current NISQ devices~\footnote{It is an interesting problem to optimize $N_{\rm{reads}}$ such that it maximizes the sensitivity of the score towards differentiating two probability distributions. This problem is left for future work.}.

For statistical robustness, we recommend as a good practice to perform $R$ repetitions of the $N_{\rm{reads}}$ measurements leading to $R$ independent estimates of the recall (each denoted as $r_i$). For estimating the precision $p$, all the samples collected should be used to robustly estimate this quantity. Using this value of $p$ one can compute $R$ independent values of the qBAS$(n,m)$ score from each of the $r_i$. These are subsequently bootstrapped to obtain a more robust average for the final reported value of qBAS$(n,m)$ (see details in Section~\ref{s:methods}).

We note that a more general performance indicator than qBAS$(n,m)$ score is indeed the KL divergence, $D_{KL}[P_{{\rm BAS}(n,m)} | P_{\boldsymbol{\theta}}]$. However, this would not be robust in terms of scalability; as $n \times m$ becomes large, it is expected that the KL divergence is frequently undefined \footnote{One may argue that the same scalability issue applies to the cost function in Eq.~\eqref{e:cost_func} used for learning. However, we can choose alternative scalable cost functions, as we show in the Supplementary Material. A comprehensive study is left for future work.}. This is true when measurements yield distributions such that $P_{\boldsymbol{\theta}}(\boldsymbol{x}^{(d)}) = 0$ for any of the $\boldsymbol{x}^{(d)}$ in BAS$(n,m)$. In all these cases, the qBAS score can still be computed and the number of measurements $N_{\rm{reads}}$ necessary for obtaining a robust estimate continues to remain relatively small to be practical for intermediate size $n \times m$.

\subsection*{Experiments}\label{s:results}
We investigated three different examples, namely, GHZ state preparation, coherent thermal state preparation, and BAS$(2,2)$. We implemented each example of DDQCL using both numerical simulations and experiments. We explored the parameter space of DDQCL by varying the qubit-to-qubit connectivity topology (see top-right panel of Figure~\ref{f:bas}) and the number of layers. We evaluated the performance of each instance by using the KL divergence from the circuit probability distribution in the computational basis to the target probability distribution. While explicitly computing the KL divergence is generally intractable, due to its demanding resource requirement as the size of instances to consider increases, we were able to compute this quantity explicitly for all cases considered in this paper.

\subsubsection*{GHZ state preparation}
To test the capabilities of DDQCL, we started with the preparation of GHZ states, also known as ``cat states''~\cite{GHZ1990}. Besides their importance in quantum information, the choice is motivated by their simple description and by the availability of many studies about their preparation (see e.g. Refs.~\cite{Monz2011,Rocchetto2017,Ozaeta2017}). From the DDQCL perspective, we explored whether it is possible to learn any of the known recipes for GHZ state preparation starting only from classical data. More specifically, the input data consists of samples from a distribution corresponding to the two desired computational basis states; $P_0=0.5$ for the state $\ket{0 \ldots 0}$ and $P_1= 0.5$ for the state $\ket{1 \ldots 1}$. Using a layer of single qubit rotations followed by an entangling layer with the all-to-all topology, DDQCL yielded many degenerate preparations of GHZ-like states differing only by a relative phase.

\begin{figure}[t]
\centering
\includegraphics[width=\columnwidth]{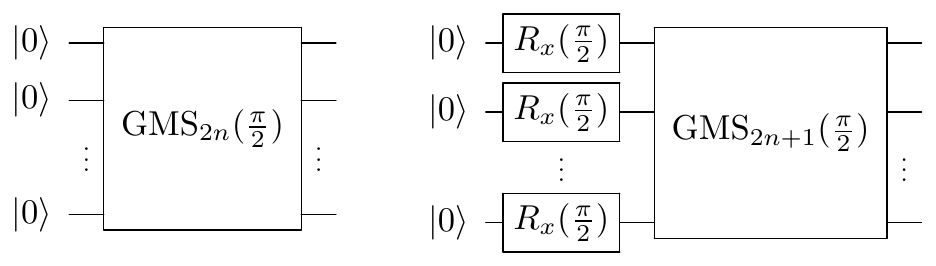}
\caption{{\it GHZ-like state preparation assisted by DDQCL.} Left (right) panel shows a recipe obtained by a human expert assisted by DDQCL for the even (odd) cat state preparation. $R_x$ stands for the single qubit rotation about the $x$ axis. GMS stands for a global M{\o}lmer-S{\o}rensen gate~\cite{Maslov2017} acting on all the $N = 2n$ ($N = 2n+1$) qubits and is equivalent to the application of local $XX$ gates to all $N(N-1)/2$ pairs of qubits. All parameters attained values very close to $\frac{\pi}{2}$. The human expert rounded the parameters to some precision and found these patterns.}
\label{f:ghz} 
\end{figure}

In particular, we first ran particle swarm optimization on 25 random initializations for 3, 4, 5 and 6-qubit instances. Then, a human expert inspected the best set of parameters learned for each size and after rounding the parameters to some precision, spotted a clear pattern. Instances of 3 and 5 qubits yielded a recipe, while instances of 4 and 6 qubit yielded another recipe. The recipes obtained are summarized in Figure~\ref{f:ghz} and were verified for larger number of qubits, both odd and even. Indeed, DDQCL successfully reproduced the recipes previously used on ion trap quantum computers~\cite{Blatt2011}, and, to the best of our knowledge, they correspond to the most compact and efficient protocols for GHZ state preparation using $XX$ gates (see Figure~\ref{f:ghz}). Another commonly used approach consists of cascading entangling gates, with alternations of single qubit rotations~\cite{Ozaeta2017}. DDQCL produced approximate recipes of this kind in some of the test cases for 3 and 4 qubits when using a single entangling layer with chain topology.

It is interesting to note that in DDQCL all the parameters are learned independently and not constrained to be the same. As shown in Figure~\ref{f:ghz}, the learning process unveiled that these converge to the same value. This is not necessarily the case for the other data sets considered below. We also note that the simulations assumed noiseless hardware, making the analysis of parameters easier for the human expert. It would be much more difficult to analyse parameters found with noisy hardware, as DDQCL can learn to compensate certain types of noise, e.g., systematic parameter offsets, in non-trivial ways. The upside is that learning can be successful even in the presence of such systematic errors.

Finally, it is reassuring that DDQCL obtained circuits for cat state preparation starting from samples of a classical distribution. While this target distribution could have been modeled by a zero-temperature ferromagnet, there apparently was no other way for our circuit to reproduce such a distribution, if not by preparing a GHZ-like state. One can obtain more general solutions by allowing DDQCL to prepare mixed states. For example, consider a circuit acting on both the main qubit register and an additional ancilla register. By tracing out the ancilla register, e.g. ignoring it during measurement, the main register can implement a mixed state and can be trained to simulate a zero-temperature ferromagnet. Another example which does not resort to an ancilla register is to use decoherence as a mechanism to prepare mixed states that explain the data.

\begin{figure*}
\centering
\includegraphics[width=.98\textwidth]{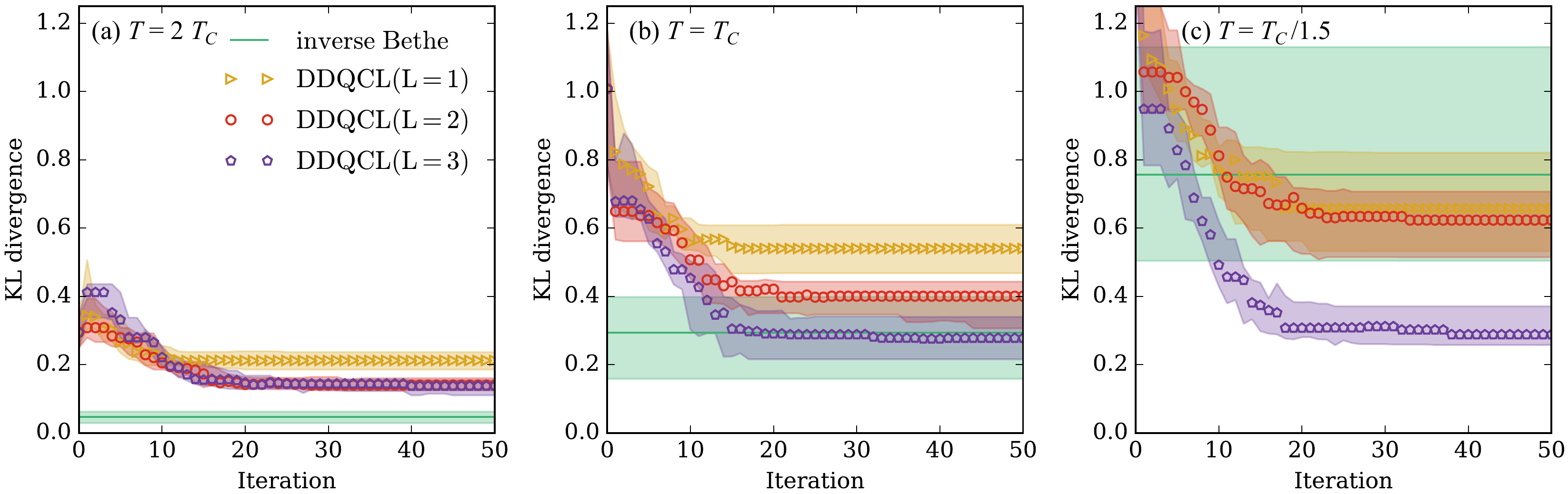}
\caption{{\it DDQCL preparation of coherent thermal states.} We generated $25$ random instances of size $N=5$ and varied the difficulty of the learning task by decreasing the temperature in $T \in \{2T_{\rm{c}}, T_{\rm{c}}, T_{\rm{c}}\slash 1.5\}$ where $T_{\rm{c}}$ is the reference temperature (see Section~\ref{s:methods} for details). The model is a quantum circuit with five qubits and an all-to-all qubit connectivity for the entangling layer. We show the bootstrapped median and $90\%$ confidence interval over the distribution of medians of the KL divergence of DDQCL as learning progresses for $50$ iterations. (a) When $T>T_{\rm{c}}$, the learning task is easy and shallow quantum circuits such as $L=1$ (yellow triangles) and $L=2$ (red circles) perform very well. (b) When $T \approx T_{\rm{c}}$, a gap in performance between circuits of different depth becomes evident. (c) When $T < T_{\rm{c}}$, the learning task becomes hard and deeper circuits perform much better than shallow ones. We also report results for the inverse Bethe approximation, which does not actually prepare a state, but produces a classical model in closed-form. The classical model so obtained (green band) is excellent for the easy task in (a), matches the best quantum model in (b), and underperforms for the hard task in (c).}
\label{f:thermal5Q}
\end{figure*}

\subsubsection*{Coherent thermal states}
Thermal states play an important role in statistical physics, quantum information, and machine learning. Using DDQCL, we trained quantum circuits with different number of layers $L \in \{1,2,3\}$ and using all-to-all topology, to approximate a target Boltzmann distribution. In particular, we considered data sets sampled from the Boltzmann distribution of $25$ synthetic instances with $N=5$ qubits. By decreasing the temperature $T$ of the target distribution, we can increase the difficulty of the learning task. Figure~\ref{f:thermal5Q} shows the bootstrapped median and $90\%$ confidence interval over the distribution of medians of the KL divergence during learning. Deeper circuits such as $L=3$ (purple pentagons) consistently outperformed shallower circuits such as $L=2$ (red circles) and $L=1$ (yellow triangles). This became more evident as we went from easy learning tasks (Figure~\ref{f:thermal5Q}~(a)) to hard learning tasks (Figure~\ref{f:thermal5Q}~(c)). Results for instances of $N=6$ qubits are shown in Figure~\ref{f:thermal6Q} in the Supplementary Material.

To assess how well DDQCL performs on the generative task, we compared DDQCL to the inverse Bethe approximation~\cite{Ricci-Tersenghi-JSTAT-2012} (see also Eqs. $(3.21)$ and $(3.22)$ in Ref.~\cite{Mastromatteo-PhD-2013}), a classical closed-form approach widely used in statistical physics to solve the inverse Ising problem. As shown in Figure~\ref{f:thermal5Q}, the inverse Bethe approximation (green bar) performed extremely well in the easy task (a), matched the $L=3$ quantum circuit in the intermediate task (b), and underperformed on the difficult task (c). The latter observation comes from the fact that the median performance of the inverse Bethe approximation has very large confidence intervals. We emphasize that this is not a form of quantum supremacy as the two methods are fundamentally different. DDQCL prepares a quantum state without the assumption of an underlying Boltzmann distribution, while the inverse Bethe approximation infers the parameters with such assumption. Furthermore, the error in the inverse Bethe approximation is expected to go to zero with system size, and only above the reference temperature $T_c$ (see Section~\ref{s:methods} for details). Thus, it is not surprising that we obtained bad performance in Figure~\ref{f:thermal5Q} (c) with the inverse Bethe approximation. Other classical methods based on machine learning and Markov Chain Monte Carlo, such as the Boltzmann machine~\cite{ackley1985learning}, could achieve higher accuracy by requiring more computational resources than the inverse Bethe approximation used here. A thorough comparison is beyond the scope of this work.

\subsubsection*{BAS(2,2)}
For the purposes of benchmarking and measuring the power of NISQ devices with DDQCL, it is insufficient to have an easy-to-generate target data set; we also require the data set to represent a useful quantum state in quantum computing, while simultaneously proving to be sufficiently challenging for the quantum computer to generate. Because of the importance of entanglement in quantum information processing, we considered the entanglement entropy averaged over all two-qubit subsets~\cite{higuchi2000how} as a proxy measure of a specific quantum state's usefulness for benchmarking purposes. We start by noting that the four-qubit cat state, whose rich entangled nature makes it ideal for studying decoherence and decay of quantum information~\cite{Monz2011,Ozaeta2017}, has entanglement entropy $S_{GHZ} = 1$. Now consider states that encode BAS$(2,2)$ in the computational basis. The minimum value of entanglement entropy that any such state can have is $S_{BAS(2,2)}=1.25163$. Furthermore, the maximum value that a quantum representation of BAS$(2,2)$ can reach is $S_{BAS(2,2)} = 1.79248$, which happens to be the maximum entanglement entropy known for any four-qubit state~\cite{higuchi2000how} (see also Figure~\ref{f:entropy} and the corresponding Section in the Supplementary Material). 

We also note that one of the quantum representations of BAS$(2,2)$ found by DDQCL reached a remarkable value of $S_{BAS(2,2)} = 1.69989$ (see Figure~\ref{f:detail_exp_vs_sim} in the Supplementary Material). This shows the power of our framework, in that DDQCL is capable of handling useful quantum states that are rich in entanglement. This is an important observation, since we know, based on our empirical results, that (i) single layer circuits with no entangling gates severely underperform in producing the output state probability distribution that is close to the target data set, and (ii) when inspecting the parameters learned for circuits with all-to-all topology with $L=2$ layers, we found that most of the $XX$ gates reached their maximum entangling setting.

We now discuss results for the qBAS$(2,2)$ score, which we computed experimentally and theoretically in order to compare the entangling topologies sketched in the top right panel of Figure~\ref{f:bas} and for different number of layers. The process consists of two steps; first, DDQCL is used to encode BAS$(2,2)$ in the wave function of the quantum state. Second, the best circuits, i.e. those with lowest cost, are compared using the qBAS$(2,2)$ score.

The bottom left and bottom central panels in Figure~\ref{f:bas} show the bootstrapped median of the KL divergence and $90\%$ confidence interval over $25$ random initializations of DDQCL \textit{in silico}. The all-to-all topology (orange circles) always outperforms sparse topologies (red stars and purple squares). However, deeper circuits do not always provide significant improvements, as it is the case for all-to-all $L=4$ (dark green circles) versus all-to-all $L=2$ (orange circles). A possible explanation is that, when going from two to four layers, we approximately double the number of parameters, and particle swarm optimization struggles to find enhanced local optima. Another plausible explanation is that for this small data set, the all-to-all circuits with $L=2$ are already close to optimal performance (we show supporting evidence in Figure~\ref{f:detail_exp_vs_sim} in the Supplementary Material).

As the best performing circuits to compare using the qBAS score, we chose all-to-all $L=2$ and star $L\in \{2,4\}$ circuits. While they represent very different approximate solutions to the same problem, they may be compared with the help of qBAS$(2,2)$ score. For each setting, we computed $25$ scores from batches of size $N_{\rm{reads}} = 15$ samples, as described in Section~\ref{s:qbas}. The bottom right panel in Figure~\ref{f:bas} shows the bootstrapped mean qBAS$(2,2)$ score and $95\%$ confidence interval for simulations (green bars) and experiments on the ion trap quantum computer hosted at University of Maryland (pink bars). The score is sensitive to the depth of the circuit as shown by the performance improvement of $L=4$ compared to $L=2$ in the star topology. Note that the theoretical improvement for using $L=4$ is larger than that observed experimentally in the ion trap. This is because the quantum computer accumulated errors while executing the deeper circuit. The score is also sensitive to the choice of topology as shown by the drop in performance of star compared to all-to-all when the same number of layers $L=2$ is used. 

Although we compared circuits implemented on the same ion trap hardware, the score may be used to compare different device generations or even completely different architectures (e.g. superconductor-based versus atomic-based). Similarly, one may use the score to compare classical resources of the hybrid system (e.g. different optimizers).

\section{Discussion}\label{s:outlook}
Data is an essential ingredient of any machine learning task. In this work, we presented a data-driven quantum circuit learning algorithm (DDQCL) as a framework that can assist in the characterization of NISQ devices and to implement simple generative models. The success of this approach is evidenced by the results on three different data sets.

To summarize, first, we learned a GHZ state preparation recipe for an ion trap quantum computer. Minimal intervention by a human expert allowed to generalize the recipe to any number of qubits. This is not an example of compilation, but rather an illustration of how simple classical probability distributions can guide the synthesis of interesting non-trivial quantum states. Depending on the level or type of noise in the system, the same algorithm could lead to a different circuit fulfilling the same probability distribution as that of the data. The message here is that machine learning can teach us that ``there is more than one way to skin a cat (state)".

Second, we trained circuits to prepare approximations of thermal states. This illustrates the power of Born machines~\cite{Cheng2017} to approximate Boltzmann machines~\cite{ackley1985learning} when the data require thermal-like features.

Finally, tapping into the real power of near-term quantum devices and approximate algorithms implementable on them, we designed a task-specific performance estimator based on a canonical data set. The bars and stripes data is easy to generate, visualize and verify classically, while modeling it still requires significant quantum resources in the form of entanglement. Errors in the device will affect this single performance measure, the qBAS score, which can be used to compare different device generations, or completely different architectures. The qBAS score can also be used to benchmark the typical performance of optimizers used in hybrid quantum-classical systems. Selecting the method and optimizing the hyper-parameters can be a daunting task and is a key challenge towards a successful implementation as the number of qubits increases. Therefore, having this unique metric for benchmarking could help reduce the complexity of this fine-tuning stage. The score can be computed in any of the NISQ architectures available to date.

DDQCL is a modular framework and its performance will ultimately depend on the choices made for such modules. In this article we explored the impact of circuit layout and cost function, while subsequent work has analysed other modules and suggested extensions to the algorithm. In Ref.~\cite{liu2018differentiable} the authors trained Born machines using a differentiable cost function and exploiting gradient calculations proposed in Ref.~\cite{mitarai2018quantum}. In Refs.~\cite{Zhu2018,leyton2019robust} the authors compared several optimizers, and in Ref.~\cite{Hamilton2018} the authors focused on the impact of hardware noise. The expressive power of shallow circuits was investigated in Ref.~\cite{du2018expressive} and it was shown to outweigh that of some classes of artificial neural networks. Finally, recent work has shown successfull implementation of DDQCL on the IBM Q 20 Tokyo processor~\cite{Hamilton2018}, on a five-qubit ion-trap hosted at University of Maryland~\cite{Zhu2018}, and on the Rigetti 16Q Aspen-1 processor~\cite{leyton2019robust}.

It is left to future work to demonstrate more realistic machine learning by allowing more flexible models and employing regularization. At a finite and fixed low circuit depth, the power of the generative model can be enhanced by including ancilla qubits, in analogy to the role of hidden units in probabilistic graphical models. Regularization can be included in the cost function. In this paper, we used the negative log-likelihood which quickly becomes expensive to estimate as the size of the system increases. We reported preliminary results on alternative cost-functions that overcome the caveat and still produce satisfactory results. Layer-wise pre-training of the quantum circuit inspired by deep learning~\cite{Bengio-Book} could initialize parameters to near-optimal locations in the cost landscape. Finally, DDQCL could be generalized to learn quantum distributions or states, assuming experimental data coming from quantum experiments, e.g. quantum measurements beyond the computational basis. We think these are the most promising directions to be explored in future work.

Our approach has the bidirectional capability of using NISQ devices for machine learning, and machine learning for the characterization of NISQ devices. We hope the ideas presented here contribute to the development of further concrete metrics to help guide the architectural hardware design, while tapping into the computational power of NISQ devices.

\section{Methods}~\label{s:methods}
\subsection*{Simulation of quantum circuits \textit{in silico}}~\label{ss:simdetails}
We simulated quantum circuits using the \texttt{QuTiP2}~\cite{qutip2} Python library and implemented the constraints dictated by the ion trap experimental setting. In the current experimental setup, we can perform arbitrary single qubit rotations and M{\o}lmer-S{\o}rensen $XX$ entangling gates involving any two qubits. We used only these gates hence avoiding the need of further compilation. For the simulations \textit{in silico}, we also assume perfect gate fidelities and error-free measurements.

In the ion trap setting, the implementation of single qubit rotations $R_z$  is very convenient. Therefore, we perform arbitrary single qubit operations relying on the decomposition ${U^{(l)}_i = R_z(\theta^{(l,3)}_i) R_x(\theta^{(l,2)}_i) R_z(\theta^{(l,1)}_i)}$, where $l$ is the layer number, $i$ is the qubit index, and ${\theta^{(l,k)}_i \in [-\pi, +\pi]}$ are Euler angles. The rotations are then expressed as exponentials of Pauli operators ${R_z(\theta^{(l,\cdot)}_i) = \exp ( -\frac{i}{2} \theta^{(l,\cdot)}_i \sigma^z_i )}$ and ${R_x(\theta^{(l,\cdot)}_i) = \exp ( -\frac{i}{2} \theta^{(l,\cdot)}_i \sigma^x_i )}$. 

Because we execute circuits always starting from the $\ket{0 \cdots 0}$ state, the first set of $R_z$ rotations would have no effect and, therefore, is not needed. When an odd number of layers is used, a similar exception occurs in the last layer. There, the last set of $R_z$ rotations would only add a phase that becomes irrelevant when taking the amplitude squared required for the Born machine. In other words, we can slightly reduce the number of parameter without changing the expressive power of the circuit. Every other layer of arbitrary single qubit operations would in general require $3N$ parameters, where $N$ is the number of qubits. By using an alternative decomposition, namely ${U = R_x R_z R_x}$, we could apply commutation rules with $XX$ gates and obtain a reduction to $2N$ parameters in all odd layers. We decided not to do the former step for two reasons. First, there is no effective reduction of the number of parameters for experiments up to $L=5$ layers considered here. Second, in the ion trap quantum computer used here~\cite{Debnath2016} it is experimentally convenient to use a larger number of $R_z$ rather than $R_x$ rotations.
 
For the case of the entangling gates, we use the notation ${U^{(l)}_{ij} = XX(\theta^{(l)}_{ij})}$, which in exponential form reads as ${XX(\theta^{(l)}_{ij}) = \exp ( -\frac{i}{2} \theta^{(l)}_{ij} \sigma^x_i \sigma^x_j )}$. Recalling that states that differ by a global phase are indistinguishable, a direct computation shows that the tunable parameters can be taken as ${\theta^{(l)}_{ij}\in[-\pi, +\pi]}$. Also, there is no need to set up an order for these gates within an entangling layer as they commute with one another. 

The total number of parameters per entangling layer depends on the chosen topology: \textit{all} is a fully-connected graph and has $N(N-1)\slash 2$ parameters, \textit{chain} is a one-dimensional nearest neighbor graph with $N-1$ parameters, and \textit{star} is a star-shaped graph with $N-1$ parameters. The top right panel of Figure~\ref{f:bas} shows a graphical representation of these topologies for the case of $N=4$ qubits.

When executing DDQCL, we always estimated the required quantities from $1000$ measurements in the computational basis.

\subsection*{Gradient-free optimization}
Once the number of layers and topology of entangling gates is fixed, the quantum circuits described above provide a template; by adjusting the parameters we can implement a small subset of the unitaries that are in principle allowed in the Hilbert space. The variational approach aims at finding the best parameters by minimizing a cost function. For all our tests, we choose to minimize a clipped version of the negative log-likelihood. 

We use a global-best particle swarm optimization algorithm~\cite{shi1998modified} implemented in the \texttt{PySwarms}~\cite{pyswarms2017} Python library. A `particle' corresponds to a candidate solution circuit; the position of a particle is a point $\boldsymbol{\theta}$ in parameter space, the velocity is a vector determining how to update the position in parameter space. Position and velocity of all the particles are initialized at random and updated at each iteration following the schema shown in Figure~\ref{f:ddvqa}. There are three hyper-parameters controlling the swarm dynamics: a cognition coefficient $c_1$, a social coefficient $c_2$ and an inertia coefficient $w$. After testing a grid of values, we chose to use a constant value of $0.5$ for all three hyper-parameters, which we found to work well for our purpose. To avoid large jumps in parameter space, we further restrict position updates in each dimension to a maximum magnitude of $\pi$. 

Finally, we set the number of particles to twice the number of parameters of the circuit. This is a conservative value compared to previous work~\cite{kandala2017hardware}, also because of the large number of parameters in the circuits explored here.

\subsection*{Data sets details}
We worked with three synthetic data sets: zero-temperature ferromagnet, random thermal, and bars and stripes (BAS). In all our numerical experiments we use $1000$ data points sampled exactly from these distributions. 

\subsubsection*{GHZ state preparation}
The zero-temperature ferromagnet distribution is equivalent to assigning $1\slash 2$ probability to both $|0\dots0\rangle$ and $|1\dots1\rangle$ states of the computational basis. This distribution can be easily prepared as a mixed state, but our study uses pure states prepared by the circuit. The only way to reproduce the zero-temperature ferromagnet distribution in our setting is to implement a unitary transformation that prepares a GHZ-like state. 

\subsubsection*{Thermal states}
A thermal data set in $N$ dimensions is generated by exact sampling realizations of $\bx \in \{-1,+1\}^N$ from the distribution ${P(\bx) = Z^{-1} \exp((\sum_{ij} J_{ij} x_i x_j + \sum_{i} h_i x_i) T^{-1}})$ where $Z$ is the normalization constant, $J_{ij}$ and $h_i$ are random coefficients sampled from a normal distribution with zero mean and $\sqrt N$ standard deviation, and $T$ is the temperature. In the large system-size limit, a phase transition is expected at $T_{\rm{c}} \approx 1$. Although this is not true for the small-sized systems considered here, we take this value as a reference temperature. In our study, we vary ${T \in \{ 2T_{\rm{c}}, T_{\rm{c}}, T_{\rm{c}} \slash 1.5\}}$ in order to generate increasingly complex instances.

\subsubsection*{Bars and Stripes}
BAS~\cite{MacKay-book-2002} is a canonical machine learning data set for testing generative models. It consists of $n \times m$ pixel pictures generated by setting each row (or column) to either black ($-1$) or white ($+1$), at random. In generative modeling, a handful of patterns are input to the algorithm and the target is to train a model to capture correlations in the data. Assuming a successful training, the model can reconstruct and generate previously unseen patterns from partial or corrupted data. On the other hand, if we provide the algorithm with {\it all} the patterns we are interested in and aim to a model that generates only those, this would amount to an associative memory. Although both tasks can be done with our DDQCL pipeline, for the qBAS$(n,m)$ score we focus on the latter task. 

We now determine the number of patterns and provide an easy identification of the bitstring belonging to the BAS$(n,m)$ class. For the total count of the number of patterns, we first count the number of single stripes, double stripes, etc. that can fit into the $n$ rows. This number is the sum of binomial coefficients $\sum_{k=0}^n \binom{n}{k} = 2^n$. The same expression holds for the number of patterns with bars that can be placed in the $m$ columns, that is $2^m$. Note that empty (all-white) and full (all-black) patterns are counted in both the bars and the stripes. Therefore, we obtain the total count for the BAS patterns by subtracting the two extra patterns from this double count:
\be
N_{{\rm BAS}(n,m)} = 2^n + 2^m - 2 .
\ee
In the main text, we use the BAS data set to design a task-specific performance indicator for hybrid quantum-classical systems. Table~\ref{t:BASnm} shows the requirements for some values of $n$ and $m$.

\begin{table}[!htbp] 
\centering {
\begin{tabular}{cccc}
    \hline
    $(n,m)$ & $N_{\rm{qubits}}$ & $N_{{\rm BAS}(n,m)}$  &  $N_{\rm{reads}}$   \\
    \hline
    (2,2) & 4 & 6 & 15 \\
    (2,3) & 6 & 10 & 30  \\
    (3,3) & 9 & 14 & 46 \\ 
    (4,4) & 16 & 30 & 120 \\ 
    (7,7) & 49 & 254 & 1554 \\   
    (8,8) & 64 & 510 & 3475 \\      
    (10,10) & 100 & 2046  & 16780 \\          
    \hline
\end{tabular} 
\caption{Example of experimental requirements for near-term quantum computers with up to $100$ qubits. As described in the main text, $N_{\rm{reads}}$ is the number of readouts required for every estimation of the qBAS score.}
\label{t:BASnm}
}
\end{table}

\subsection*{Bootstrapping analysis}~\label{s:boot}
To obtain error bars for the KL divergence, we used the following procedure. DDQCL was always executed 25 times with random initialization of the parameters. From the 25 repetitions, we sampled 10,000 data sets of size 25 with replacement and computed the median KL divergence for each. From the distribution of 10,000 medians, we computed the median and obtained error bars from the 5-th and 95-th percentiles as the lower and upper limits, respectively, accounting for a $90\%$ confidence interval.

For the case of qBAS score, we did the following bootstrap analysis. qBAS score was always computed 25 times from batches of samples $N_{\rm{reads}}$. From the 25 repetitions, we sampled 10,000 data sets of size 25 with replacement and computed the mean for each. From the distribution of 10,000 means, we computed mean and obtained error bars from two standard deviations, accounting for a $95\%$ confidence interval.

\section*{Data availability} 
All data needed to evaluate the conclusions are available from the corresponding author upon request.

\section*{Acknowledgements}
M.B. was supported by the UK Engineering and Physical Sciences Research Council (EPSRC) and by Cambridge Quantum Computing Limited (CQCL). The authors are very grateful to Prof. Christopher Monroe and his team at the University of Maryland (UMD) for their support in running the experiments presented here. Special thanks to N.M. Linke, C. Figgatt, K.A. Landsman, and D. Zhu for useful discussions and for running the several experiments used to test and validate the pipeline of this work, and the ones presented in the Results section. The authors acknowledge E. Edwards from the communications/publicity division at the Joint Quantum Institute at UMD for the rendering of the ion-trap graphic used in Figure~\ref{f:bas}, and would like to thank J. Realpe-Gomez, A.M. Wilson, and G. Paz-Silva for useful discussions and feedback on an early version of this manuscript. The authors would like to thank J.I. Latorre for pointing out Ref.~\cite{higuchi2000how}.

\section*{Author contributions}
M.B. and A.P-O designed the generative model and the learning algorithm. M.B., Y.N., and A.P-O designed the data sets and the experiments. M.B. and V.L-O wrote the code and performed the \textit{in silico} experiments. D.G-P wrote code for the analysis and figures. A.P-O designed the qBAS score. O.P. performed the analysis related to entanglement entropies. All authors analyzed the experimental results and contributed to the final version of the manuscript.

\section*{Competing interests} 
The authors declare no competing interests.

\widetext

\vspace{25pt}

\raggedbottom

\begin{center}
\textbf{\large Supplementary Material for ``A generative modeling approach for benchmarking and training shallow quantum circuits''}
\end{center}

\subsection*{Comparison of cost functions}~\label{s:costs}
In realistic machine learning scenarios, we typically do not have access to the complete target probability distribution, nor to  that obtained from the output state of a quantum circuit. Hence, we need to compare the two distributions at the level of histograms and using a finite number of samples and measurements. Here we compare three cost functions via simulations \textit{in silico}.

First, we defined the \textit{clipped negative log-likelihood} as
\be\label{e:nll}
\mathcal{C}_{nll} (\boldsymbol{\theta}) = - \frac{1}{D}\sum_{d=1}^{D} \ln (\max (\epsilon, P_{\boldsymbol{\theta}}(\mathbf{x}^{(d)}))) ,
\ee
where probabilities are estimated from samples and $\epsilon>0$ is a small number that avoids an infinite cost when $P_{\boldsymbol{\theta}}(\mathbf{x}^{(d)})=0$. This may happen if for example entanglement in the circuit prevents us from measuring configuration $\mathbf{x}^{(d)}$. Moreover, when the number of variables $N$ is large, all except few configurations will ever be measured due to the finite number of samples. Note that by re-normalizing all the probabilities after the clipping, we could interpret this variant as a Laplace additive smoothing. All the experiments in the main text are carried out using the clipped negative log-likelihood with $\epsilon=10^{-8}$.

Second, we defined the \textit{earth mover's distance}~\cite{rubner2000earth} as
\be
\mathcal{C}_{emd} (\boldsymbol{\theta}) = \min_{F} \langle d(\mathbf{x}, \mathbf{y}) \rangle_{F} ,
\ee
where $F(\mathbf{x},\mathbf{y})$ is a joint probability distribution such that $\sum_{\mathbf{y}} F(\mathbf{x},\mathbf{y}) = P_{\mathcal{D}}(\mathbf{x})$ and $\sum_{\mathbf{x}} F(\mathbf{x},\mathbf{y}) = P_{\boldsymbol{\theta}}(\mathbf{y})$, that is, its marginals correspond to the data and circuit distributions, respectively. Intuitively, this is the minimum cost of turning one histogram into the other where the ground metric $d(\mathbf{x},\mathbf{y})$ specifies the cost of transporting a single unit from $\mathbf{x}$ to $\mathbf{y}$. We chose $d(\mathbf{x},\mathbf{y})$ to be the Hamming distance between strings $\mathbf{x}\in\{-1, +1\}^N$ and $\mathbf{y}\in\{-1, +1\}^N$. Since we normalize histograms to sum up to one, the Earth Mover's Distance is equivalent to the $1$-st Wasserstein distance~\cite{levina2001earth}. In our simulations, we use the \texttt{PyEMD} Python library for fast computation of the earth moving distance~\cite{pele2009}.

Third, we defined the \textit{moment matching} as
\be
\begin{split}
\mathcal{C}_{mm} (\boldsymbol{\theta}) &= \frac{1}{N}\sum_{i}^{N} ( \langle x_i \rangle_{P_{\mathcal{D}}} - \langle x_i \rangle_{P_{\boldsymbol{\theta}}})^2 + \frac{2}{N(N-1)}\sum_{i > j}^{N} ( \langle x_i x_j \rangle_{P_{\mathcal{D}}} - \langle x_i x_j \rangle_{P_{\boldsymbol{\theta}}})^2 ,
\end{split}
\ee
where the expectation values $P_\mathcal{D}$ and $P_{\boldsymbol{\theta}}$ are taken with respect to data and circuit distributions, respectively. This cost function can be generalized to include moments beyond the second as well as using different positive exponents for the error.

\begin{figure}[H]
\centering
\includegraphics[width=.38\textwidth]{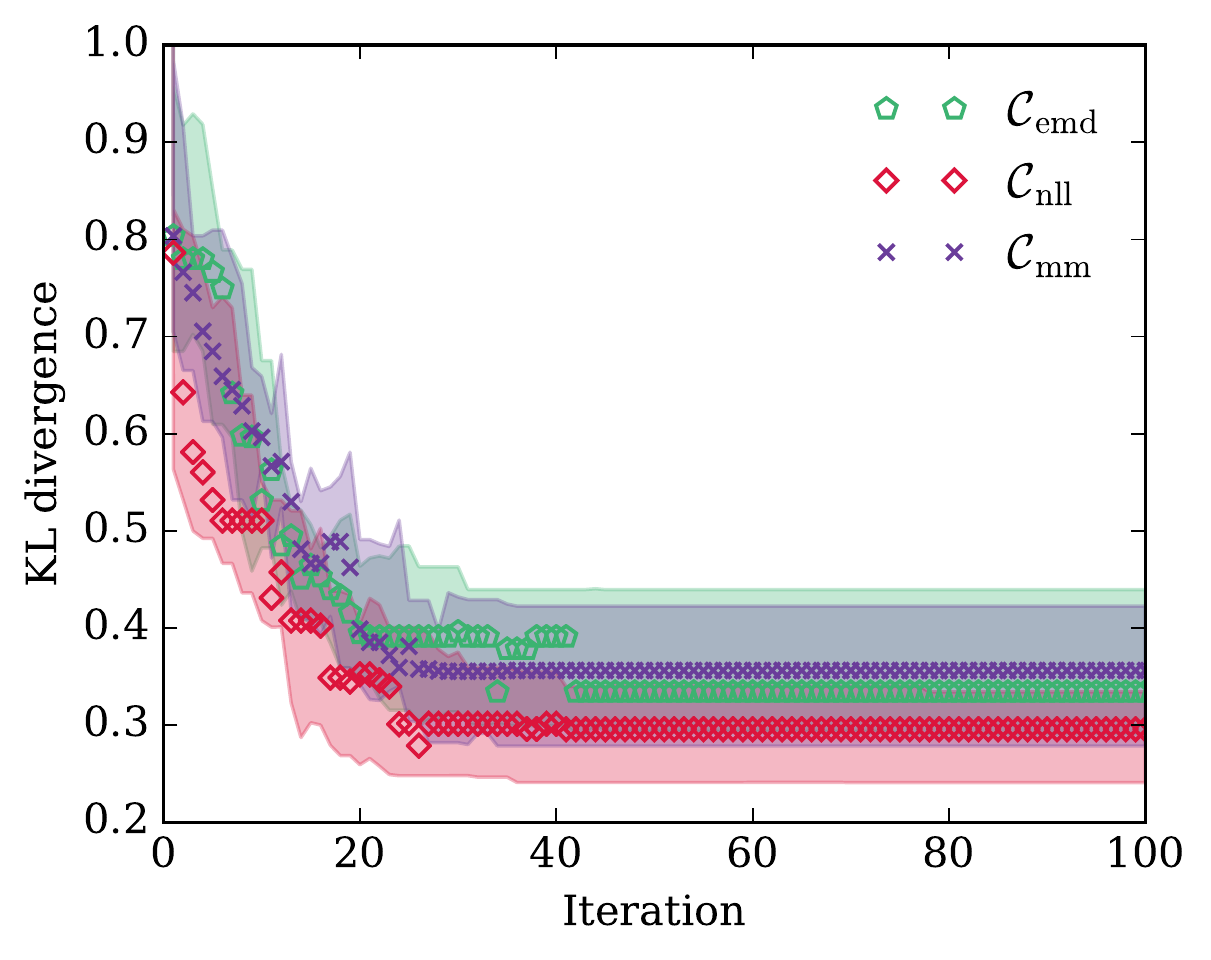}
\caption{Bootstrapped median KL divergence and $90\%$ confidence interval of circuits learned by minimizing different cost functions. Both the moment matching ($\mathcal{C}_{mm}$) and earth mover's distance ($\mathcal{C}_{emd}$) closely track the clipped negative log-likelihood ($\mathcal{C}_{nll}$) used in the main text.}~\label{f:costs}
\end{figure}

We compared the cost functions on the task of learning thermal states of size $N=5$, with $L=3$ layers and \textit{all} topology. Figure~\ref{f:costs} shows the bootstrapped median KL divergence on $25$ realizations and $90\%$ confidence interval as learning progress for $100$ iterations. The fact that $\mathcal{C}_{nll}$ (red diamonds) outperforms other cost functions does not come as a surprise; minimization of the negative log-likelihood is indeed directly related to minimization of the KL divergence. However, we expect the performance of DDQCL based on this cost function to degrade quickly as the size of the problem increases. In realistic applications, the relevant probabilities in Eq.~\eqref{e:nll}, i.e. those associated with the data, are a vanishing fraction of the $2^N$ probabilities. Moreover, they need to be estimated from a finite number of measurements. The earth mover's distance $\mathcal{C}_{emd}$ (green pentagons) performs well, but it suffers from similar scalability issues. Fast algorithms for the computation of this distance may struggle when the number of bins in the histogram increases exponentially as in our case. However, it is reassuring to see that alternative cost functions with no relation to the KL divergence can still produce satisfactory results. Surprisingly, the moment matching $\mathcal{C}_{mm}$ (purple crosses) closely tracks the other cost functions while retaining computational efficiency. In fact, even though a large number of samples may be needed to obtain low-variance estimates for the moments, only $\mathcal{O}(N^2)$ terms are computed at each iteration. We expect this cost function to be a good heuristic for DDQCL on large systems. 

\subsection*{Approximate preparation of coherent thermal states for \texorpdfstring{$N=6$}{N=6}}~\label{s:thermal6Q}
\begin{figure}[H]
\vspace{-.75cm}
\centering
\includegraphics[width=.99\textwidth]{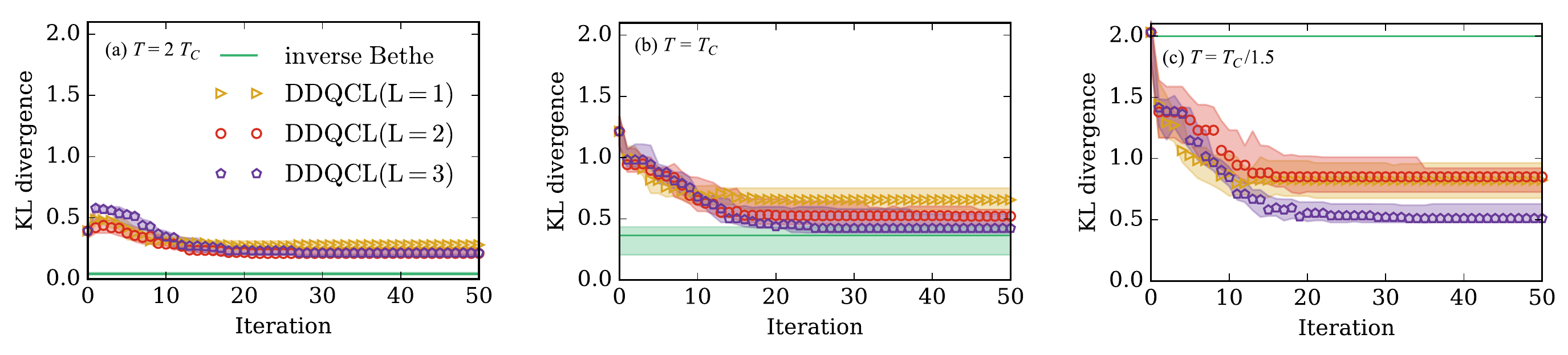}
\caption{DDQCL preparation of approximate coherent thermal states for the case of six qubits, different circuit depths $L \in \{1,2,3\}$, and different temperatures ${T \in \{ 2T_{\rm{c}}, T_{\rm{c}}, T_{\rm{c}} \slash 1.5\}}$. In these simulations, an all-to-all topology was used for the entangling gate layer ($l=2$). We report the bootstrapped median and $90\%$ confidence interval. For the low temperature case shown in panel (c), the inverse Bethe approximation converged only in 7 out of 25 instances. Hence, no median value was extracted. We plotted a KL divergence of 2.0 as a reference, but it is clear that for all those instances DDQCL outperformed the inverse Bethe approximation.}
\label{f:thermal6Q}
\end{figure}

\subsection*{Details for theoretical and experimental results}~\label{s:exp_vs_sim}
\vspace{-.75cm}
\begin{figure}[H]
\centering
\includegraphics[width=.99\textwidth]{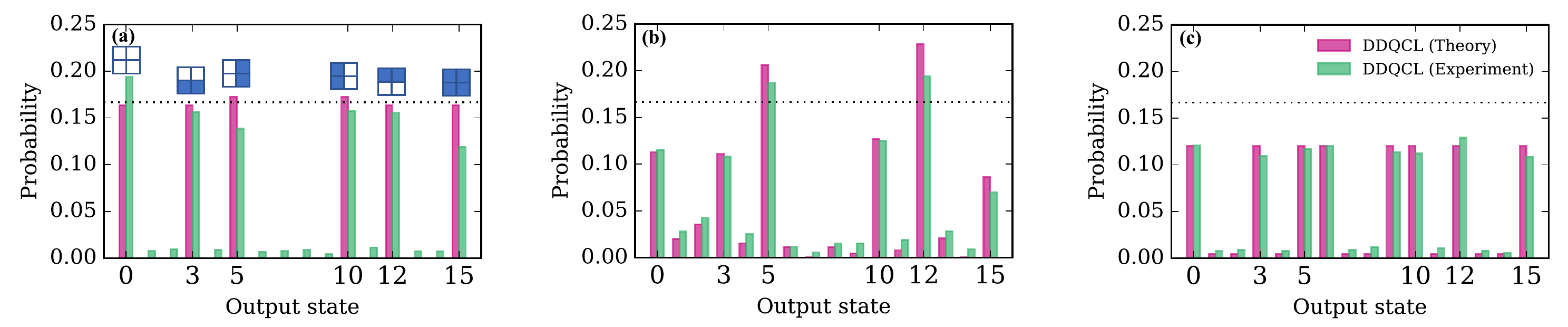}
\caption{Detailed comparison at the level of output states between simulation of circuits and the respective experimental implementations in the ion trap quantum computer. These are the best circuits in terms of KL divergence that were obtained by DDQCL for BAS$(2,2)$ under three different setting: (a) all-to-all with $L=2$ layers, (b) star with $L=4$, and (c) star with $L=2$. The theoretical state obtained in (a) is close to optimal and attains an entanglement entropy averaged over all two-qubit subsets of $S_{BAS(2,2)}=1.69989$. Circuit diagrams for (a-c) are shown in Figure~\ref{f:circuits}.}
\label{f:detail_exp_vs_sim}
\end{figure}

\begin{figure}[H]
\begin{tabular}{ll}
\large (a) & \raisebox{-.5\height}{\includegraphics[height=80pt]{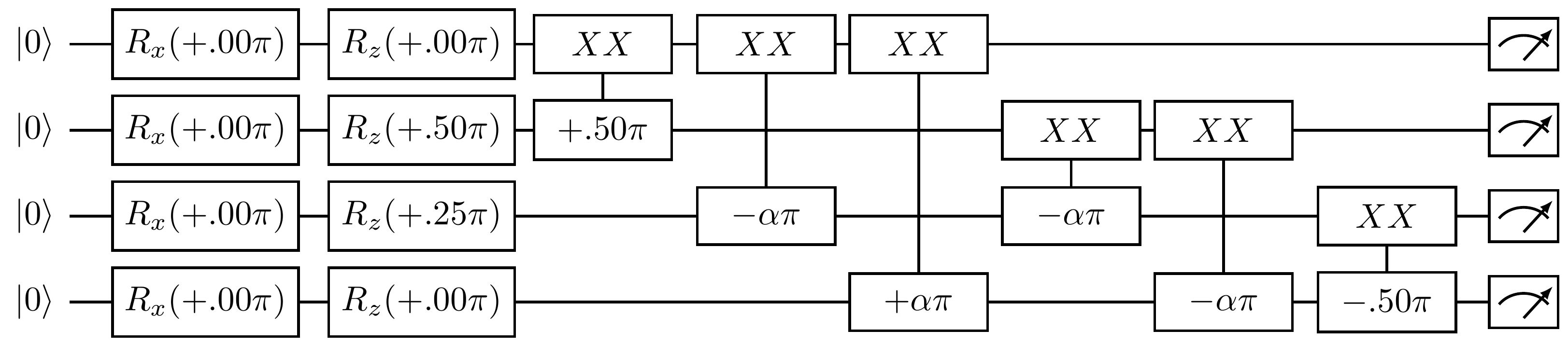}}\\
&\\&\\
\large (b) & \raisebox{-.5\height}{\includegraphics[height=80pt]{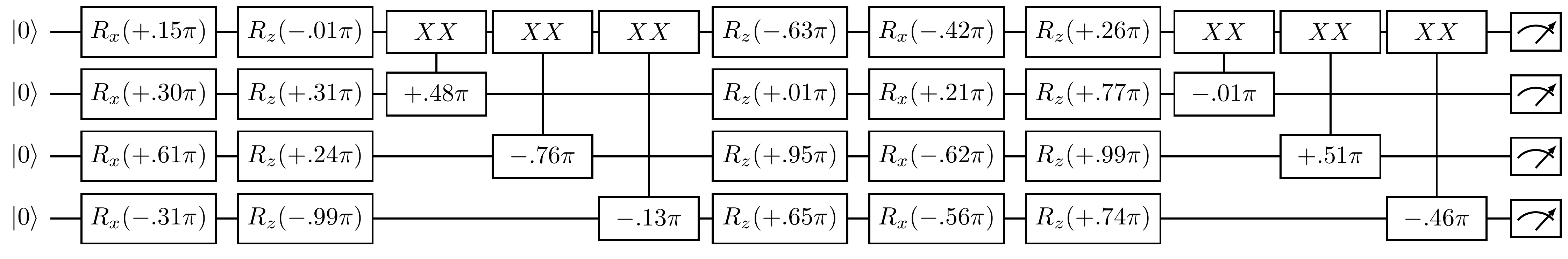}}\\
&\\&\\
\large (c) & \raisebox{-.5\height}{\includegraphics[height=80pt]{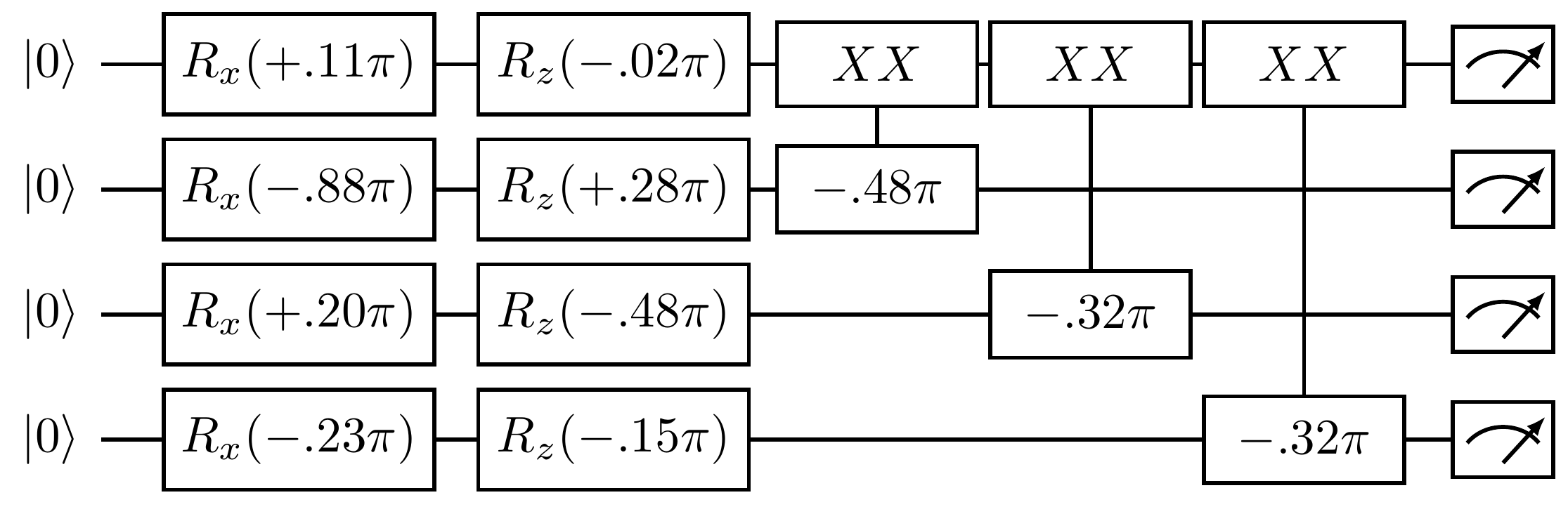}}\\
\end{tabular}
\caption{Circuit diagrams for the analytical solution and for the best circuits found by DDQCL under three different qubit-to-qubit connectivity topologies. (a) the all-to-all circuit with $L=2$ layers can achieve zero KL divergence for BAS$(2,2)$ by setting $\alpha = \pi^{-1} \arctan(2^{-1/2})$. All single-qubit rotations can be set to zero. DDQCL found an almost optimal solution with $\alpha=0.2$ and two non-zero $R_z$ rotations. These $R_z$ gates act as the identity on the $\ket{0000}$ state. (b) the star circuit with $L=4$. (c) the star circuit with $L=2$. The qBAS scores for (a-c) are shown in Figure~\ref{f:bas} in the Main Text.}
\label{f:circuits}
\end{figure}

\subsection*{Entanglement entropy of BAS(2,2)}~\label{s:entropy}
The measure of entanglement entropy used in this work is the average von Neumann entropy over all 2-qubit subsets~\cite{higuchi2000how}. Consider a 4-qubit pure state ${\rho = \ket{\psi} \bra{\psi}}$ and label the four qubits as A, B, C, and D. Then, the entropy can be computed as
\be
S_\psi = -\frac{1}{3} \Big [ \tr( \rho_{AB} \log_2 \rho_{AB} ) + \tr( \rho_{AC} \log_2 \rho_{AC} ) + \tr( \rho_{AD} \log_2 \rho_{AD} ) \Big ] ,
\ee
where $\rho_{XY}$ is the reduced density matrix for the subset $XY$. As an example, the 4-qubit cat state has an entanglement entropy of $S_{GHZ}=1$. 
Now consider a pure state encoding the uniform probability distribution over the BAS$(2,2)$ data set in the computational basis
\be
\ket{BAS(2,2)} = \frac{1}{\sqrt{6}} \left ( e^{i u_1} \ket{0000} + e^{i u_2} \ket{0011} + e^{i u_3} \ket{0101} + e^{i u_4} \ket{1010} + e^{i u_5} \ket{1100} + \ket{1111} \right ).
\ee

\begin{figure}[H]
\centering
\includegraphics[width=.44\textwidth]{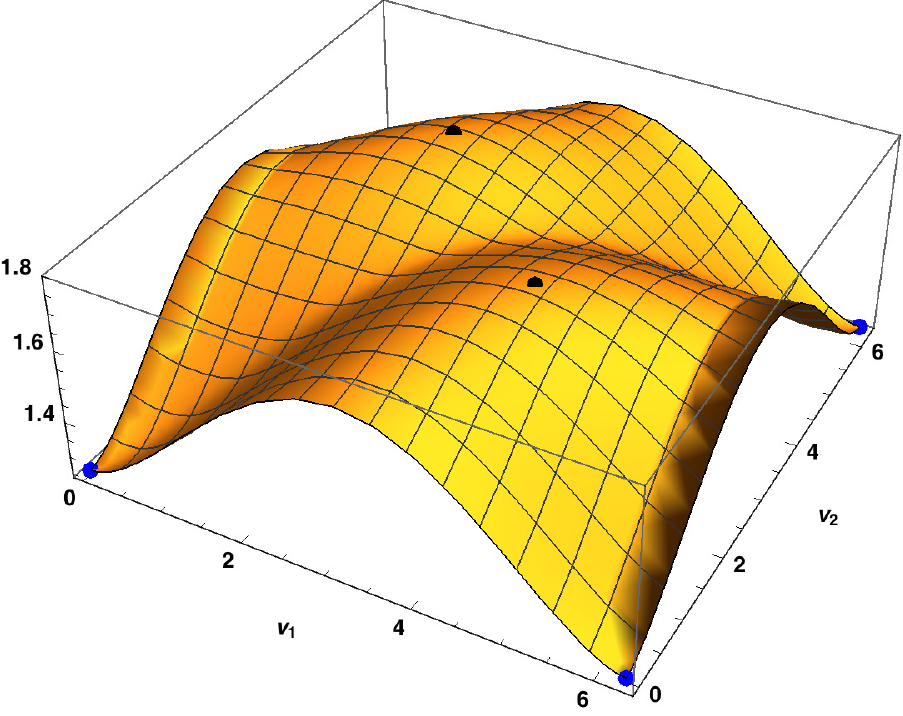}
\caption{Entanglement entropy $S_{BAS(2,2)}$ as a function of variables $v_1$ and $v_2$. Points in the domain represent states that encode the BAS$(2,2)$ data set in the computational basis. The maximum entropy (black dots) attained is $1.79248$ which coincides with the maximum entanglement entropy known for any 4-qubit state~\cite{higuchi2000how}.}
\label{f:entropy}
\end{figure}

A direct computation shows that the entropy of this state is
\be
\begin{split}
S_{BAS(2,2)} = -\tfrac{1}{9} \Big [ 
& \tfrac{2}{\ln(2)} \sqrt{\tfrac{\cos \left(u_2-u_3-u_4+u_5\right)+1}{2}} \tanh^{-1}\left(\sqrt{\tfrac{\cos \left(u_2-u_3-u_4+u_5\right)+1}{2}}\right)\\
& + \left(\cos \left(\tfrac{u_1-u_3-u_4}{2} \right)+1\right) \log_2 \left(\tfrac{2}{3} \cos^2\left(\tfrac{u_1-u_3-u_4}{4} \right)\right)
 + \left(\cos \left(\tfrac{u_1-u_2-u_5}{2} \right)+1\right) \log_2 \left(\tfrac{2}{3} \cos^2\left(\tfrac{u_1-u_2-u_5}{4} \right)\right)\\
& + \log_2 \left(4 + 2 \sqrt{2} \sqrt{\cos \left(u_2-u_3-u_4+u_5\right)+1}\right)
 + \log_2 \left(4 - 2 \sqrt{2} \sqrt{\cos \left(u_2-u_3-u_4+u_5\right)+1}\right)\\
& - \left(\cos \left(\tfrac{u_1-u_3-u_4}{2} \right)-1\right) \log_2 \left(\tfrac{2}{3} \sin^2\left(\tfrac{u_1-u_3-u_4}{4} \right)\right)
 - \left(\cos \left(\tfrac{u_1-u_2-u_5}{2} \right)-1\right) \log_2 \left(\tfrac{2}{3} \sin^2\left(\tfrac{u_1-u_2-u_5}{4} \right)\right)\\
& - \log_2 (31104) \Big ] .
\end{split}
\ee
Defining new variables $v_1=u_2-u_3-u_4+u_5$ and $v_2 =u_1-u_3-u_4$, the expression above reduces to
\be
\begin{split}
S_{BAS(2,2)} = -\tfrac{1}{9} \Big [ 
& \tfrac{2}{\ln(2)} \sqrt{ \cos^2 \left( \tfrac{v_1}{2} \right)} \tanh^{-1}\left(\sqrt{\cos^2 \left( \tfrac{v_1}{2} \right)}\right)\\
& + 2\cos^2 \left(\tfrac{v_2}{4} \right) \log_2 \left(\tfrac{2}{3} \cos^2\left(\tfrac{v_2}{4} \right)\right)
 + 2\cos^2 \left(\tfrac{v_2-v_1}{4} \right) \log_2 \left(\tfrac{2}{3} \cos^2\left(\tfrac{v_2 - v_1}{4} \right)\right)\\
& + \log_2 \left(4 + 2 \sqrt{2} \sqrt{\cos \left(v_1\right)+1}\right)
 + \log_2 \left(4 - 2 \sqrt{2} \sqrt{\cos \left(v_1\right)+1}\right)\\
& + 2\sin^2 \left(\tfrac{v_2}{4} \right) \log_2 \left(\tfrac{2}{3} \sin^2\left(\tfrac{v_2}{4} \right)\right)
 + 2\sin^2 \left(\tfrac{v_2-v_1}{4} \right) \log_2 \left(\tfrac{2}{3} \sin^2\left(\tfrac{v_2-v_1}{4} \right)\right)\\
& - \log_2 (31104) \Big ] .
\end{split}
\ee
In Figure~\ref{f:entropy} we graphically show the entropy $S_{BAS(2,2)}$ as a function of the new variables $v_1$ and $v_2$. Such a function has extrema
\be
\begin{split}
\min S_{BAS(2,2)} &= \frac{1}{3} \log_2 \left ( \frac{27}{2} \right ) \approx 1.25163 ,\\
\max S_{BAS(2,2)} &= \frac{1}{2} \log_2(12) \approx 1.79248 .
\end{split}
\ee
For the minimum value, $v_1 = v_2 = 0$, which can be obtained setting $u_1 = \cdots = u_5 = 0$. For the maximum value, $v_1 = 4 \pi /3$ and $v_2 = 2 \pi /3$, which can be obtained setting $u_1 = u_2 = u_3 = 0$ and $u_4 = -u_5 = 2 \pi/3$. Interestingly, the maximum of $S_{BAS(2,2)}$ happens to coincide with the maximum entanglement entropy known for any 4-qubit state~\cite{higuchi2000how}.

\end{document}